\documentclass[12pt]{iopart}

\usepackage{iopams}
\usepackage[usenames,dvipsnames]{xcolor}
\usepackage[square,sort&compress,numbers,comma]{natbib}
\usepackage{graphicx}
  \expandafter\let\csname equation*\endcsname\relax
  \expandafter\let\csname endequation*\endcsname\relax
\usepackage{amsmath}
\usepackage{subfig}
\usepackage{empheq}
\usepackage{epsfig}
\usepackage{caption}

\begin{document}

\title{Plasma turbulence simulations with X-points using the flux-coordinate independent approach}

\author{F.~Hariri$^1$, P.~Hill$^2$, M.~Ottaviani$^2$ and Y.~Sarazin$^2$}

\address{$^1$\'{E}cole Polytechnique F\'{e}d\'{e}rale de Lausanne (EPFL), Centre de Recherches en Physique des Plasmas (CRPP), CH-1015 Lausanne, Switzerland}
\address{$^2$CEA, IRFM, F-13108 Saint-Paul-lez-Durance, France}

\ead{farah.hariri@epfl.ch}
\vspace{10pt}
\begin{indented}
\item[] 8 September 2014
\end{indented}

\begin{abstract}
In this work, the Flux-Coordinate Independent (FCI) approach to plasma turbulence simulations is formulated for the case of generic, static magnetic fields, including those possessing stochastic field lines. It is then demonstrated that FCI is applicable to nonlinear turbulent problems with and without X-point geometry. In particular, by means of simulations with the FENICIA code, it is shown that the standard features of ITG modes are recovered with reduced toroidal resolution. Finally, ITG turbulence under the influence of a static island is studied on the transport timescale with ITER-like parameters, showing the wide range of applicability of the method.
\end{abstract}

%

\submitto{\PPCF}
%
%

\section{Introduction}
Field-aligned coordinates are necessary to optimize plasma turbulence codes. They are widely employed in today's codes allowing the number of grid points needed to represent structures elongated along the magnetic field to be greatly reduced. The Flux-Coordinate independent (FCI) approach, not based on magnetic flux variables, has been introduced and validated in~\cite{Hariri20132419, Hariri2013FENICIA}. It was also demonstrated, for the first time, in~\cite{Hariri2013FENICIA, hariri2014flux} that FCI can efficiently deal with X-point configurations. In practice, this approach is used in the direction parallel to the magnetic field, allowing to decouple the grid of the numerical problem from the magnetic field geometry. The description in the poloidal plane does not employ magnetic coordinates and information on field lines are only needed to compute parallel derivatives. Furthermore, operators can be freely discretized using any desired numerical scheme. Employing FCI opens up the way for the study of important physical processes in complex geometries, such as turbulence and magnetohydrodynamical (MHD) instabilities in a tokamak plasma. The approach was first implemented in the framework \textbf{FENICIA}~\cite{Hariri20132419,Hariri2013FENICIA}. It was then implemented in the turbulence framework \textbf{BOUT++}~\cite{BenDudson, dudson2009bout++} and in the full-f gyrofluid code \textbf{ASELA}~\cite{HeldTTF}. Currently, it is being implemented in the 3D Braginskii solver \textbf{GBS}~\cite{ricci2012simulation} and in the 5D full-f gyrokinetic code \textbf{GYSELA}~\cite{grandgirard2007global}.\\

MHD instabilities in the plasma can lead to the spontaneous appearance of usually large scale magnetic islands. They break the simple nesting of the axisymmetric toroidal surfaces, hence the axisymmetry of the magnetic equilibrium. The development of magnetic islands is particularly deleterious to the confinement for several reasons. First of all, the pressure tends to flatten inside the island as a result of the fast parallel transport, hence reducing the overall energy content of the plasma. Secondly, several magnetic islands located at different radial positions can overlap, leading to chaotic transport over large radial distances. Somewhat related is the possibility for these modes to lead to the too fast outward transport of energetic particles such as Helium ash (alpha particles), possibly preventing them from depositing their energy into the bulk plasma before escaping the confined region. Last but not least, these islands may not saturate before reaching the boundary of the plasma leading to the sudden loss of any confinement, called a disruption. It is thus of a major concern to study of the growth and the saturation mechanisms of magnetic islands in tokamaks. These issues are expected to depend on the interplay between the island and turbulence, which also affects both pressure and current transport. Nowadays, numerical tools are used to investigate the nonlinear dynamics of magnetic islands and their interaction with microturbulence, such as in~\cite{Waelbroeck1993494, ottaviani2004multiple, ishizawa2007multi, muraglia2009nonlinear, hornsby2010nonlinear, ishizawa2013magnetic}.

The rest of this paper is organized as follows. Firstly, we formulate the FCI approach for a large class of 3D magnetic field configurations including stochastic field lines. Secondly, we show for the first time the special features of using this approach in a turbulent Ion Temperature Gradient (ITG) regime. Previous applications of FCI to sound-wave models were shown in~\cite{Hariri20132419} and in~\cite{hariri2014flux}. A convergence study in the nonlinear turbulent regime is carried out in the direction of symmetry of a cylindrical plasma. We show that, with this method, we only need a few tens of toroidal points, regardless of the toroidal mode number, provided that adequate resolution is available in the poloidal plane. Finally, we investigate the influence of a static island on the transport due to ITG turbulence with ITER-like parameters, showing how turbulence is modified. In what follows, numerical simulations are carried out using the code FENICIA developed in~\cite{Hariri20132419, Hariri2013FENICIA}.

\section{The FCI approach applied to generic 3D magnetic fields}

In Ref.~\cite{Hariri20132419, hariri2014flux}, we formulated the FCI approach by a family of coordinate transformations based on a two-dimensional poloidal flux function $\psi(x,y)$. In this section, we generalize this formulation by extending it to a larger class of magnetic field configurations with good flux surfaces where $\psi(x,y,z)$ is in fact three-dimensional. We then show an even more general formulation, based on simple integration of the magnetic field line equations, which is useful when dealing with stochastic magnetic fields. As an example, the treatment of the toroidal case using magnetic field line integration is also outlined.

\subsection{Extension to 3D configurations}

The starting point of the FCI method as developed in Ref.~\cite{Hariri20132419} is a straight magnetic configuration in a three dimensional Cartesian reference system $(x,y,z)$ such that $\mathbf{\hat{z}}$ is the direction of the magnetic axis, the main magnetic field along $z$ is constant and normalized to unity. Then the magnetic field can be written as
\begin{equation}
\mathbf{B} = \pmb\nabla \times (\psi \mathbf{\hat{z}}) + \mathbf{\hat{z}} .
\label{mag-field-2}
\end{equation}
Hereafter we assume that the flux function can depend on all three coordinates, $\psi = \psi(x,y,z)$.
The parallel derivative operator is then given by
\begin{equation}
\nabla_\parallel = -[\psi,\cdot] + \partial / \partial z
\end{equation}
As in Ref.~\cite{Hariri20132419}, we look for a family of transformations of coordinates of the form:
\begin{eqnarray}
	\left\{
	\begin{array}{ll}
	\xi^\alpha &= V^\alpha+ C^\alpha (z - z_k) \label{gen-transf1-2} \\
	s &= z - z_k 
	\end{array}
	\right.
\end{eqnarray}
where, as before, we have allowed for the partition of the whole domain in overlapping sectors centered around $z_k$ and extending up to
$z_{k \pm 1}$, with $k$ labeling a given sector. 
$V^\alpha$ and $C^\alpha$ are functions of $x,y,z$ to be determined. 
In the new coordinates the parallel derivative is given by
\begin{align}
\label{par-der-gen-transf-2}
\nabla_\parallel \xi^\alpha \, \frac{\partial}{\partial \xi^\alpha} + \frac{\partial}{\partial s}
\end{align}
In order to express the parallel derivative only in terms of $s$, $\nabla_\parallel = \partial/\partial s$, the following necessary condition must be satisfied:
\begin{equation}
\nabla_\parallel \xi^\alpha = 0
\end{equation}
This translates into conditions for $V^\alpha$ and $C^\alpha$:
\begin{eqnarray}
	\left\{
	\begin{array}{ll}
	C^\alpha &= -\nabla_\parallel V^\alpha\\
	\nabla_\parallel C^\alpha &= \nabla_\parallel (\nabla_\parallel V^\alpha) = 0 \label{cond-3D}
	\end{array}
	\right.
\end{eqnarray}
Assume now that a function $\psi^*$ exists such that 
\begin{equation}
\label{eq_for_psi_star}
\nabla_\parallel \psi^* = 0
\end{equation}
Note that in the 2D case treated in Ref.~\cite{Hariri20132419}, a solution of (\ref{eq_for_psi_star}) is $\psi$ itself. In the generic 3D case, there is no guarantee that such a function exists. When it does, we refer to this situation as the integrable case. In this instance, the system possesses good magnetic surfaces identified by given values of $\psi^*$. Examples of integrable cases are the helical configurations in slab and cylindrical geometry.

In the slab case one has

\begin{equation}
\psi = \psi(x, y - \lambda z) ,
\end{equation}
such that 
\begin{equation}
\psi^* = \psi+ \lambda x .
\end{equation}

Similarly, in the cylindrical case,
\begin{equation}
\psi = \psi(r, m\theta - n \varphi ) ,
\end{equation}
and 
\begin{equation}
\psi^* = \psi+ \frac{n}{2 m}r^2 ,
\end{equation}
a well-known expression in the context of magnetic island theory.

As in the 2D case, solutions for (\ref{cond-3D}) take the form
\begin{equation}
\label{gen-sol-V-2}
V^\alpha = f^\alpha(\psi^*) + g^\alpha(\psi^*) \,\chi,
\end{equation}
with 
$\chi$ such that $\nabla_\parallel \chi = 1$. 
We remark that the existence of such a $\chi$ is guaranteed only in the integrable case when it satisfies a $1^{st}$ order linear PDE in only two variables. 
In conclusion, the transformation can be written as
\begin{eqnarray}
	\left\{
	\begin{array}{ll}
	\xi^\alpha &= f^\alpha(\psi^*) + g^\alpha(\psi^*) \,\chi - g^\alpha(\psi^*) (z - z_k) \\
	s &= z - z_k 
	\end{array}
	\right.
\end{eqnarray}
Then, similarly  to the 2D case, if one chooses to compute the parallel derivative by finite differences, one has to find $\Delta {\mathbf x}$ corresponding to a given increment $\Delta s$ along s with $\xi^\alpha = cst$. The following finite difference equation results for the unknown increments $\Delta {\bf x}$ 
\begin{equation}
[f^\alpha(\psi^*) + g^\alpha(\psi^*) \chi]_{{\mathbf x} + \Delta {\mathbf x}, z_k + \Delta z} -
[g^\alpha(\psi^*)]_{{\mathbf x} + \Delta {\mathbf x}, z_k + \Delta z} \Delta z = 
[f^\alpha(\psi^*) + g^\alpha(\psi^*) \chi]_{\mathbf x, z_k}
\end{equation}
Solutions to these FD equations exist if the following conditions are satisfied
\begin{align}
& \psi^*({\mathbf x} + \Delta {\mathbf x}, z_k + \Delta z) =\psi^*({\mathbf x}, z_k )\\
& \chi({\mathbf x} + \Delta {\mathbf x}, z_k + \Delta z) =\chi({\bf x}, z_k) + \Delta z.
\end{align}
By considering $\Delta z$ as an independent continuous variable, and $\Delta {\mathbf x}$ a function of it,
we can derive differential equations for $\Delta {\mathbf x}$:
\begin{align}
\frac{d \Delta {\mathbf x}}{d \Delta z}\cdot\pmb\nabla_{\mathbf x} \psi^* + \partial_z \psi^* &= 0 \label{Deltaxv1}\\
\frac{d \Delta {\mathbf x}}{d \Delta z}\cdot\pmb\nabla_{\mathbf x} \chi + \partial_z \chi &= 1 \label{Deltaxv2}
\end{align}
Using the fact that $\nabla_\parallel \psi^* = 1$ and $\nabla_\parallel \chi = 1$ we can see that the increments $\Delta x$ and $\Delta y$ satisfy the field-line equations:
\begin{eqnarray}
	\frac{d \Delta x}{d \Delta z} &= [\partial_y \psi]_{{\mathbf x} + \Delta {\mathbf x}, z_k + \Delta z} \label{field-tracerv1}\\
	\frac{d \Delta y}{d \Delta z} &= -[\partial_x \psi]_{{\mathbf x} + \Delta {\mathbf x}, z_k + \Delta z} \label{field-tracerv2}
\end{eqnarray}

\subsection{Direct calculation of the parallel derivative in cylindrical and in toroidal geometry}

In the case of a straight magnetic configuration, the direct calculation of $\nabla_\parallel$ at a given point is done by considering an arc of field-line passing through a point $(x,y,z)$. Using a parametrization $x(\tau), y(\tau), z(\tau)$, the field line satisfies the following system:
\begin{align}
& \frac{d x}{d \tau} = \partial_y \psi \nonumber
\\
& \frac{d y}{d \tau} = -\partial_x \psi
\\
& \frac{d z}{d \tau} = 1 , \nonumber
\end{align}
where the parameter $\tau$ is conveniently chosen in such a way that its increments are identical to those of $z$.
For any differentiable function $f(x,y,z)$, one can compute the derivative along the field-line with respect to $\tau$ as follows:
\begin{align}
\frac{d}{d \tau} f[x(\tau), y(\tau), z(\tau)] &= \frac{d x(\tau)}{d \tau} \partial_x f + \frac{d y(\tau)}{d \tau} \partial_y f + \frac{d z(\tau)}{d \tau} \partial_z f\\
& = \partial_y \psi \, \partial_x f - \partial_x \psi \, \partial_y f + \partial_z f \\
& = (\nabla_\parallel f)_{\tau}
\end{align}
This shows that $(\nabla_\parallel f)_{z = z_k}$ can be computed by finite differences with increments $\Delta \tau = \Delta z$, 
\begin{equation}
\nabla_\parallel f = \frac{f(\tau + \Delta \tau) - f(\tau - \Delta \tau)}{2\,\Delta \tau},
\end{equation} 
where the coordinates of the end points at $\tau \pm \Delta \tau$ are obtained by a field line tracer and the value of the function at those points are computed by a suitable interpolation method.
In the toroidal case, in the usual machine
polar coordinates, $(R, \varphi, Z)$, the infinitesimal displacements are $(dR, R d\varphi, dZ)$. These must be proportional to the components of $\mathbf{B}$. Thus,
\begin{align}
& d R \propto B_R d \tau \nonumber
\\
& R d \varphi \propto B_{\varphi} d \tau 
\\
& d Z \propto B_Z d \tau \nonumber
\end{align}
where $\tau$ is a parameter for the position along the field line. 
As before, it is convenient to fix $\tau$ such that $d \varphi / d \tau =1$ (case for toroidal sectors). The following field line equations result:
\begin{align}
& \frac{d R}{d \tau} = R \frac{B_R}{B_\varphi} \nonumber
\\
& \frac{d \varphi}{d \tau} = 1 \label{toroidal_field_lines}
\\
& \frac{d Z}{d \tau} = R \frac{B_Z}{B_\varphi} \nonumber
\end{align}
 Then for any function $f(R, \varphi, Z)$ 
 \begin{align}
 \frac{d}{d \tau} f(R, \varphi, Z) &= \frac{d R}{d \tau} \partial_R f + \frac{d \varphi}{d \tau} \partial_{\varphi} f + \frac{d Z}{d \tau} \partial_Z f \nonumber
 \\
& = \frac{R}{B_\varphi} [B_R \partial_R f + \frac{B_\varphi}{R} \partial_{\varphi}f + B_Z \partial_Z f ] 
\\
& = \frac{R}{B_\varphi}(\mathbf{B}.\pmb\nabla f)\nonumber
 \end{align}
 At a given point $\varphi$, this leads to the following expression in FD form:
\begin{equation}
(\mathbf{B}.\pmb\nabla f)^\varphi = \left(\frac{B_\varphi}{R}\right)^\varphi \, \left[\frac{f(\varphi + \Delta \varphi) - f(\varphi - \Delta \varphi)}{2\,\Delta \varphi}\right]
\end{equation}  
where $f(\varphi \pm \Delta \varphi)$ corresponds to the value of $f(R,Z,\varphi)$ at the points $(R^{\pm},Z^{\pm},\varphi \pm \Delta \varphi)$ where $R^{\pm}$ and $Z^{\pm}$ are obtained by integrating the field-line equations~\eqref{toroidal_field_lines}. 

It is important to notice that this formulation, not being based on a coordinate transformation, does not require the existence of integrals of the field line equations such as the $\psi^*$. One can then conclude that the FCI approach is also applicable to configurations with stochastic field lines.  

\section{An application to ITG turbulence}\label{sec:ITGturbulence}

The aim of the present section is to highlight the special advantages underlying the use of the Flux-Coordinate Independent (FCI) approach for ITG turbulence simulations. To this end, a 3-dimensional 4-field fluid system, which models the slab branch of the ITG instability, serves as a testbed. The most complete system of equations which has been implemented in the framework FENICIA, is shown here:
\begin{equation}
\label{eq:ITG_complete}
\left\{\begin{array}{ll}
\partial_t \tilde n + {[\phi,log(n_0)] - [\phi,\rho_*^2 \nabla_\perp^2\phi]} + {C_\parallel \nabla_\parallel u_\parallel} = {D_n \nabla_\perp^2 \tilde n} \\\\
\partial_t u_\parallel + {[\phi,u]} + {C_\parallel (\frac{1}{\tau}\nabla_\parallel \tilde{n} + \nabla_\parallel \phi + \nabla_\parallel T_\parallel)} = {D_u \nabla_\perp^2 u_\parallel} \\\\
\partial_t T_\perp + {[\phi,T_\perp]} - {\chi_{\parallel \perp} \nabla_\parallel^2 T_\perp} = {D_{T_\perp}\nabla_\perp^2 T_\perp} \\\\
\partial_t T_\parallel + {[\phi,T_\parallel]} + {\frac{2}{\tau} C_\parallel \nabla_\parallel u_\parallel - \chi_{\parallel \parallel} \nabla_\parallel^2 T_\parallel}  = {D_{T_\parallel}\nabla_\perp^2 T_\parallel} \\\\
\tilde n = \phi
\end{array}
\right.
\end{equation}
The first equation stands for the continuity (or mass conservation) equation, the second one for the parallel momentum balance, and the last two for transverse and parallel heat transport equations, respectively. Notice that, in the present version of the model, the variable $T_\perp$ appears only in one equation. It acts as a \emph{passive scalar}, in the sense that it is sensitive to the turbulent field, namely $\phi$, but it does not back-react on it. This quantity can be considered as a tracer of the turbulent field. All quantities are dimensionless. $\tilde n$ is the relative perturbed ion guiding center density, $n_0$ is the equilibrium density profile, $u_\parallel$ is the ion parallel velocity normalized to the thermal speed $v_{th}$, $T_\perp$ and $T_\parallel$ the transverse and parallel ion temperatures normalized to the constant electron temperature $T_e$, and $\phi$ is the electrostatic potential normalized to $T_e/e$. Both transverse coordinates $(x,y)$ involved in the Poisson bracket are normalized to the tokamak minor radius $a$. We define two dimensionless parameters: $C_\parallel = a/R \times 1/\rho_*$ where $R$ is the tokamak major radius, $\rho_* = \rho_s / a$ is the reduced gyro-radius with $\rho_s = (mT_e)^{1/2}/eB$ being the ion sound Larmor radius. Moreover $\tau$ is the ratio of electron temperature to ion temperature $T_e/T_i$. Time is normalized to the Bohm timescale $a^2/(\rho_s c_s)$, where $c_s = (T_e/m)^{1/2}$ is the ion sound speed. The explicit expression of the parallel derivative operator $\nabla_\parallel$ depends on the magnetic field structure. In the case of a cylindrical geometry one can write
$$\nabla_\parallel = \partial_\varphi + 1/q(r) \: \partial_\theta$$
with ($r,\theta$) the polar coordinates in the poloidal plane and $q(r)$ the safety factor. Finally, $D_n$, $D_u$, $D_{T_\perp}$ and $D_{T_\parallel}$ are dissipative transverse transport coefficients which account for the weak collisional transport and ensure the damping of small scales. $\chi_{\parallel \perp}$ and $\chi_{\parallel \parallel}$ are the collisional parallel transport coefficients of transverse and parallel temperature, respectively.\\

Nonlinear turbulence is herein studied numerically by solving model~\eqref{eq:ITG_complete}, first in a cylindrical geometry. It is convenient to start with a discussion on the properties of the turbulent regime embedded in the model described by~\eqref{eq:ITG_complete}.
\begin{figure}
  \centering
	\includegraphics[width=0.6\linewidth,keepaspectratio,clip]{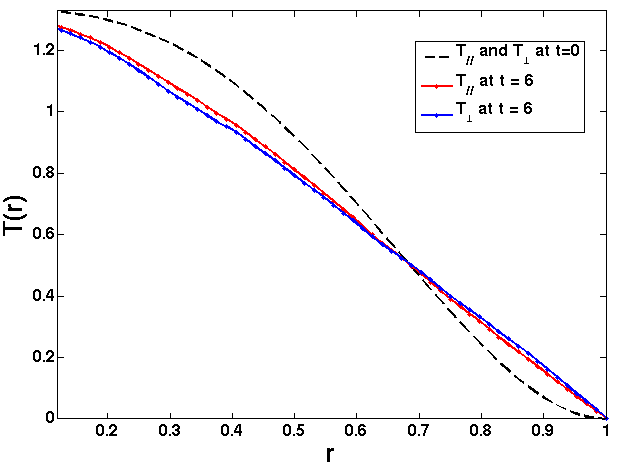}
  \caption{Parallel and perpendicular temperature profiles at $t=0$ and at the $t=6$, the time at the end of the simulation}
  \label{T_profiles}
\end{figure}
The main parameters used for the nonlinear simulations are chosen as follows: the code is run until $t=6$ with a normalized time step equal to $dt = 10^{-3}$. We recall here that time is normalized to the Bohm time scale $a^2/(\rho_sc_s)$, and lengths to the minor radius $a$. The differential operators are evaluated on a grid of size $(nx, ny, nz) = (200, 200, 20)$ with $\rho_* = 0.08$. It follows that the transverse grid increment is equal to $\rho_s/16$. The transverse dissipative coefficients and the parallel transport coefficient are set to $D_\perp = 10^{-3}$ and $\chi_\parallel = 12.5$. Periodic boundary conditions are used in the $\mathit{z}$ direction and the vector of states is set to zero outside the plasma radius. We consider a density profile expressed as $log(n_0) = -(r^2 - a^2)/2L_n$ and a $q$ profile defined by $q=1+2r^2$. The perturbed density is initialized by
$$\tilde{n} (t=0,r) = \sum g_0(r) \times \cos(m\theta  - n\varphi )$$
where
$$g_0 (r) = \exp{ \left [{- \frac{(r-r_s)^2}{r_s^2}m^2} \right ]} \times \left (\frac{r}{r_s} \right )^m \times \left (\frac{r-a}{r_s-a}\right ) ^2.$$
and the sum involves only two modes $(m,n) = (9,6)$ and $(m,n) = (10,7)$. The radial envelope of the initial modes retains a Gaussian centered on their rational surface $r_s$ (where $q(r_s)=m/n$, so that $k_\parallel(r_s) = 0$), and is chosen sufficiently narrow so as to minimize their parallel wave vector.\footnote{Indeed, a Taylor expansion of the parallel wave vector around the resonance position $r_s$ leads to: $k_\parallel(r) \simeq (r-r_s)\; dk_\parallel/dr|_{r_s} = -(r-r_s)k_\theta/L_s$, with the poloidal wave vector equals to $k_\theta=m/r_s$ and the magnetic shear length defined by $L_s=qR/s$.}
As far as the initial equilibrium profiles are concerned, parallel velocity is such that $u_\parallel(t=0,r) = 0$ and both the parallel and perpendicular temperatures are chosen to be equal at $t=0$. The initial profile for both of the temperatures is plotted in Fig.~\ref{T_profiles} (dashed line).
\begin{figure}
		\centering
		\includegraphics[width=0.8\linewidth,keepaspectratio,clip]{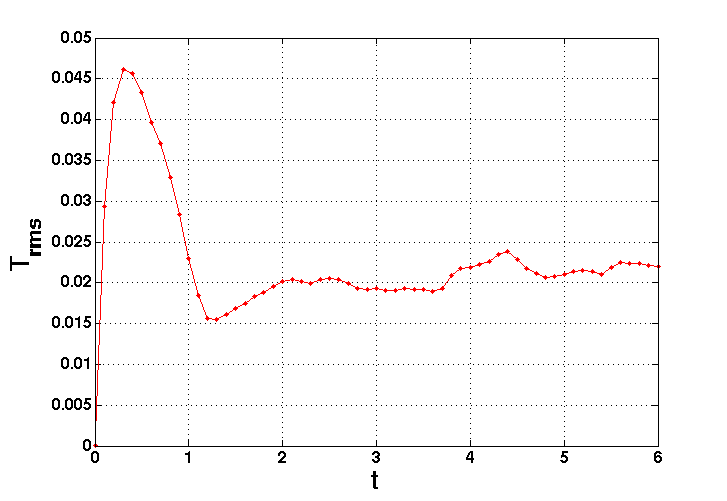}
        \caption{$T_{rms}$ as a function of time}
        \label{fig:Trms_1D}
\end{figure}
Recall that the code is global, i.e. that there is no scale separation between equilibrium and fluctuating quantities, including temperatures. Since no heat source is added to sustain the profiles so far, one expects the initial temperature profiles to relax under the action of heat turbulent transport. The result is actually visible in Fig.~\ref{T_profiles}, where the final temperature profiles are plotted. The gradient at the initial stage exceeds its critical value at the position were the perturbations were initiated (at roughly a mid-radius position on the simulation domain). Fluctuations then become unstable, and grow exponentially during the linear phase and eventually saturate as evidenced in the time evolution of the root mean square of the temperature fluctuations~Fig.~\ref{fig:Trms_1D}. The magnitude of the fluctuations exhibits an overshoot to reach a peak value at $t\sim 0.3$, which results from the dynamical balance between the linear excitation and nonlinear saturation mechanisms. The latter refer to dissipation via mode-mode coupling, which involves energy cascade towards linearly stable modes, and profile relaxation leading to the reduction of the linear growth rate. Considering the fairly large $\rho_*$ value of the reported simulation, the energy confinement time is quite small, so that the time scale for profile relaxation $-$ which is of the order of the energy confinement time for the entire profile $-$ competes with that of nonlinear energy transfer. On one hand, assuming a Bohm-like scaling for $\tau_E$ leads to $\omega_c\tau_E \sim \rho_*^{-2}$ and on the other hand, nonlinear energy transfer typically occurs on an eddy turn-over time $\omega_c\tau_{eddy} \sim (k_\theta v_E)^{-1} \sim [(k_r\rho_s)(k_\theta\rho_s)(e\phi/T)]^{-1}$. Using a mixing length type of argument, then $e\phi/T \sim \rho_*$, and $k_r\rho_s \sim k_\theta\rho_s <1$. It readily appears that, in this type of regime, both times are of the same order of magnitude: $\tau_E \sim \tau_{eddy}$. Therefore, both nonlinear mechanisms are likely to be effective in this case, and efficiently contribute to the nonlinear saturation of turbulence.  \\

\begin{figure*}
  \centering
 \hspace*{-0.5cm} \vspace*{-5mm}\subfloat[t=0]{\includegraphics[width=0.5\linewidth,keepaspectratio,clip]{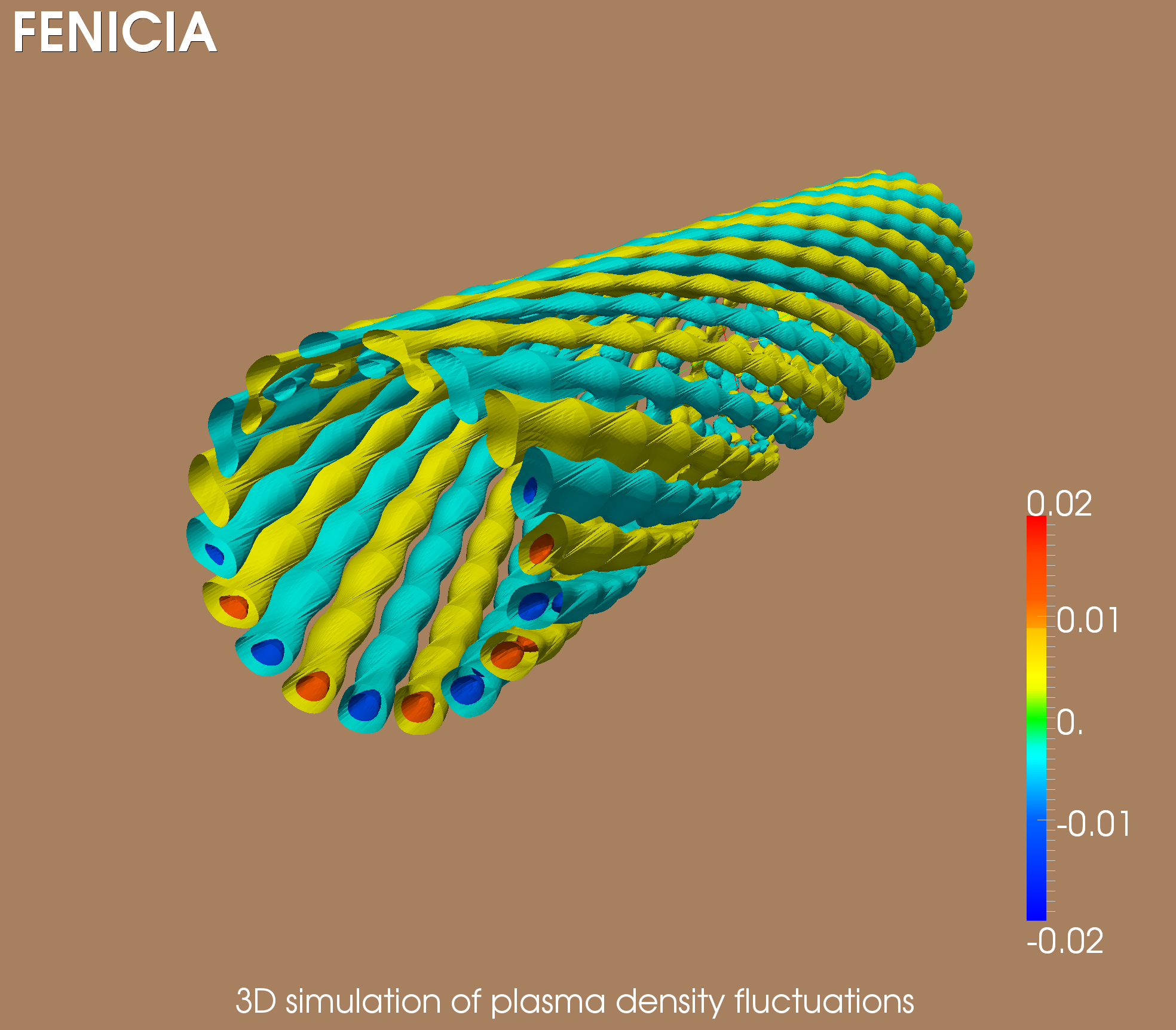}} \hspace*{3ex}
 \hspace*{-0.5cm} \subfloat[t=0.9]{\includegraphics[width=0.5\linewidth,keepaspectratio,clip]{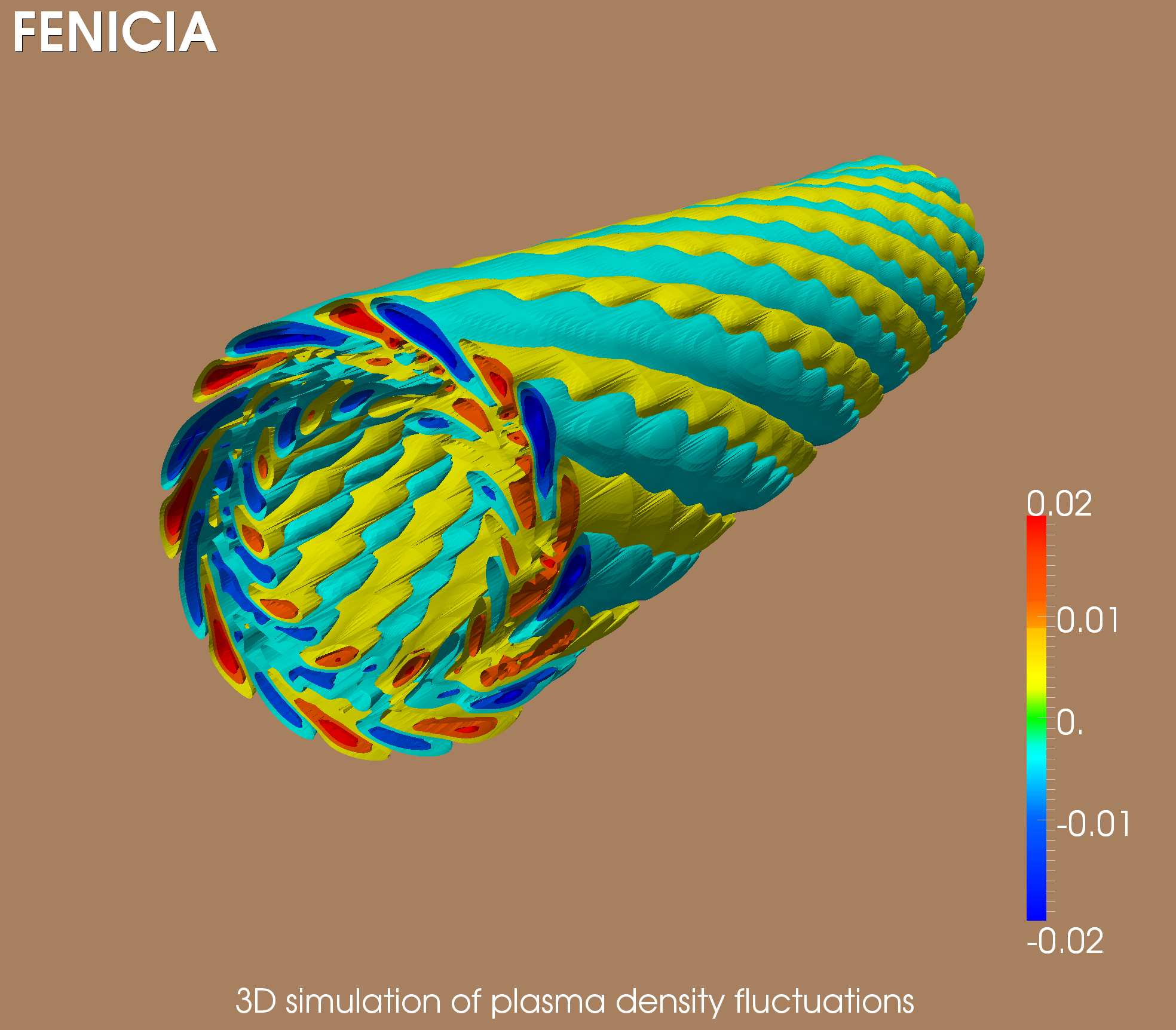}}\\ \vspace*{1ex}
 \hspace*{-0.49cm}\subfloat[t=2]{\includegraphics[width=0.5\linewidth,keepaspectratio,clip]{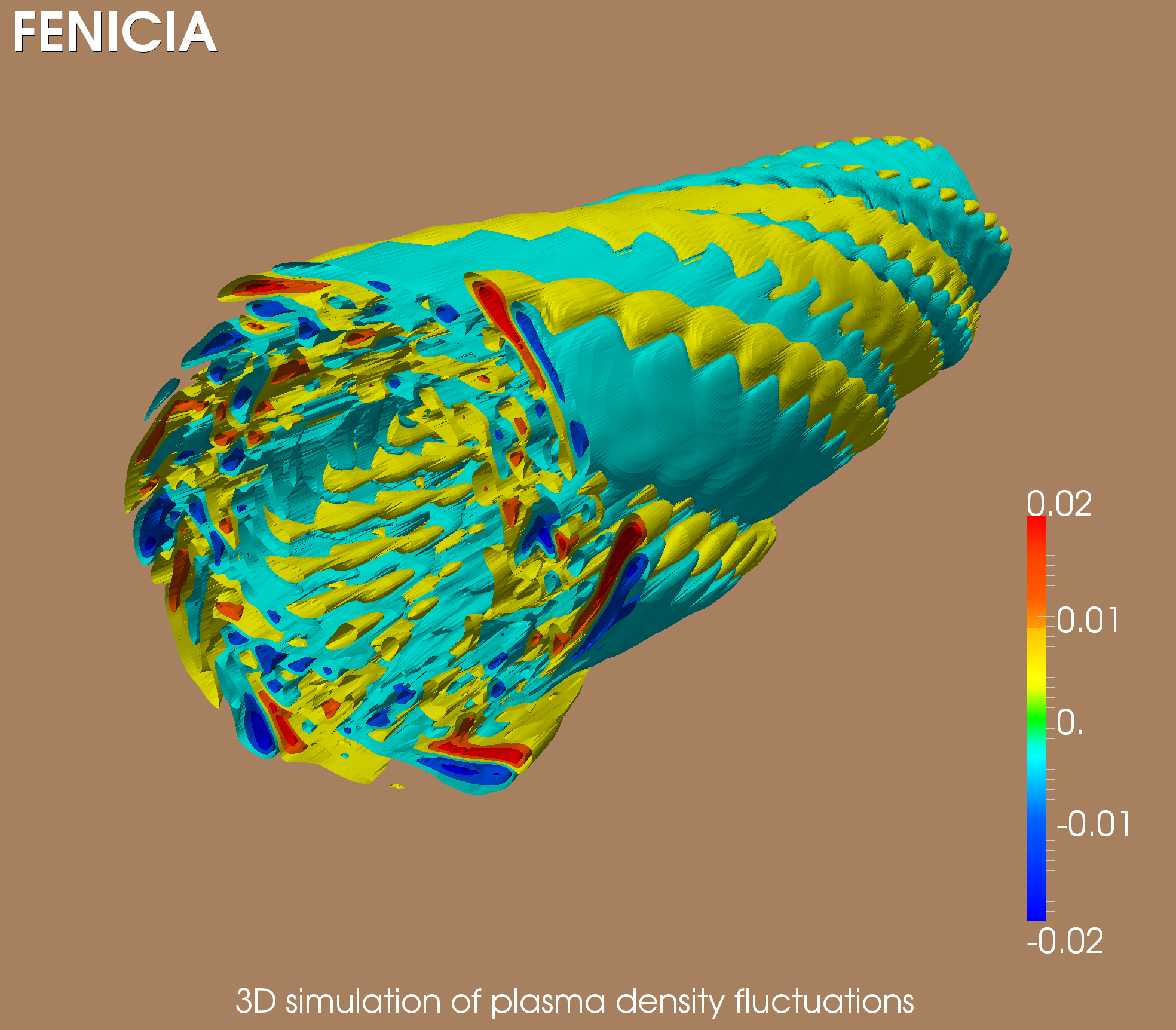}} \hspace*{3ex}
 \hspace*{-0.49cm} \subfloat[t=3]{\includegraphics[width=0.5\linewidth,keepaspectratio,clip]{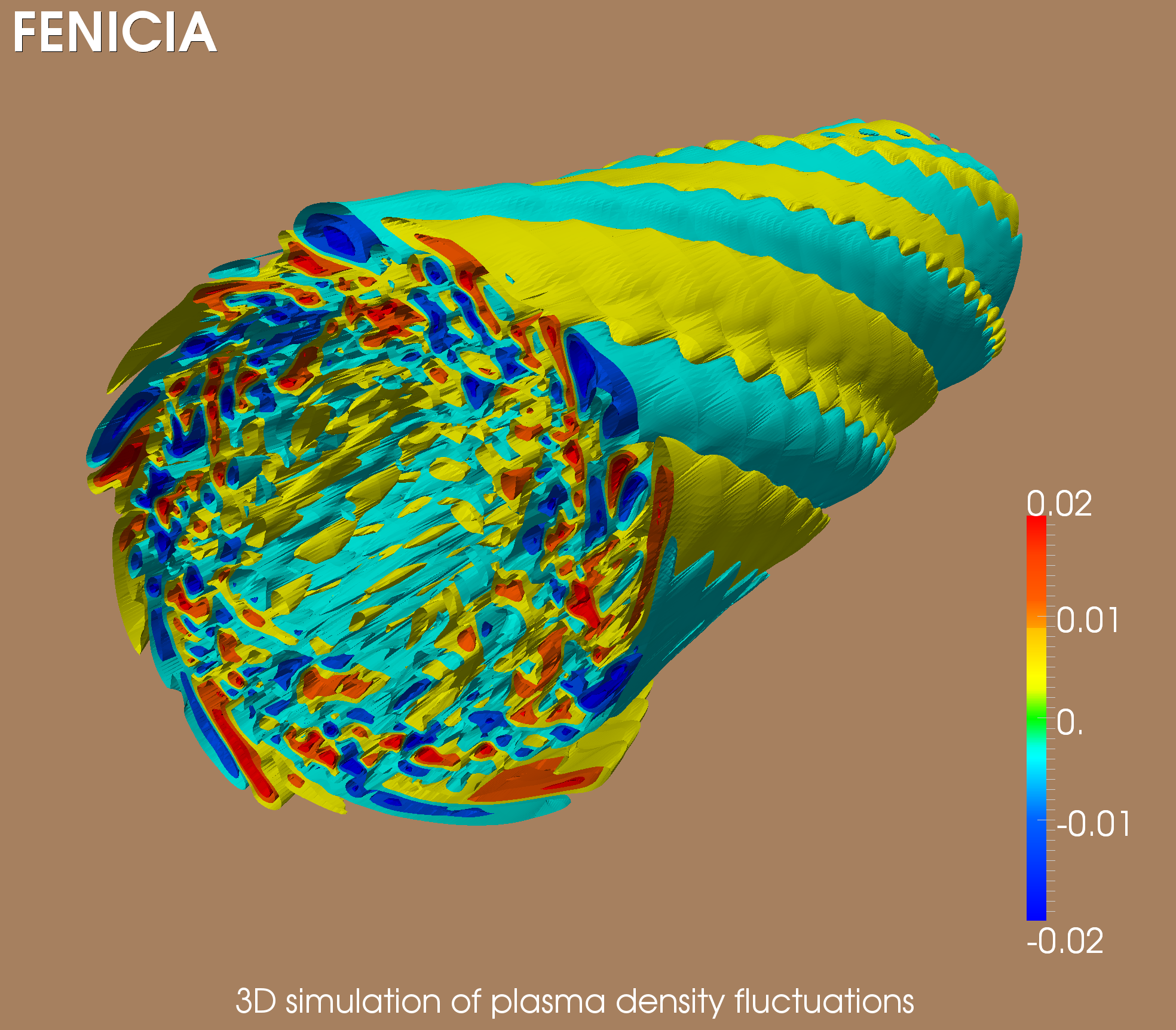}}\\
 \hspace*{-0.5cm} \subfloat[t=4]{\includegraphics[width=0.5\linewidth,keepaspectratio,clip]{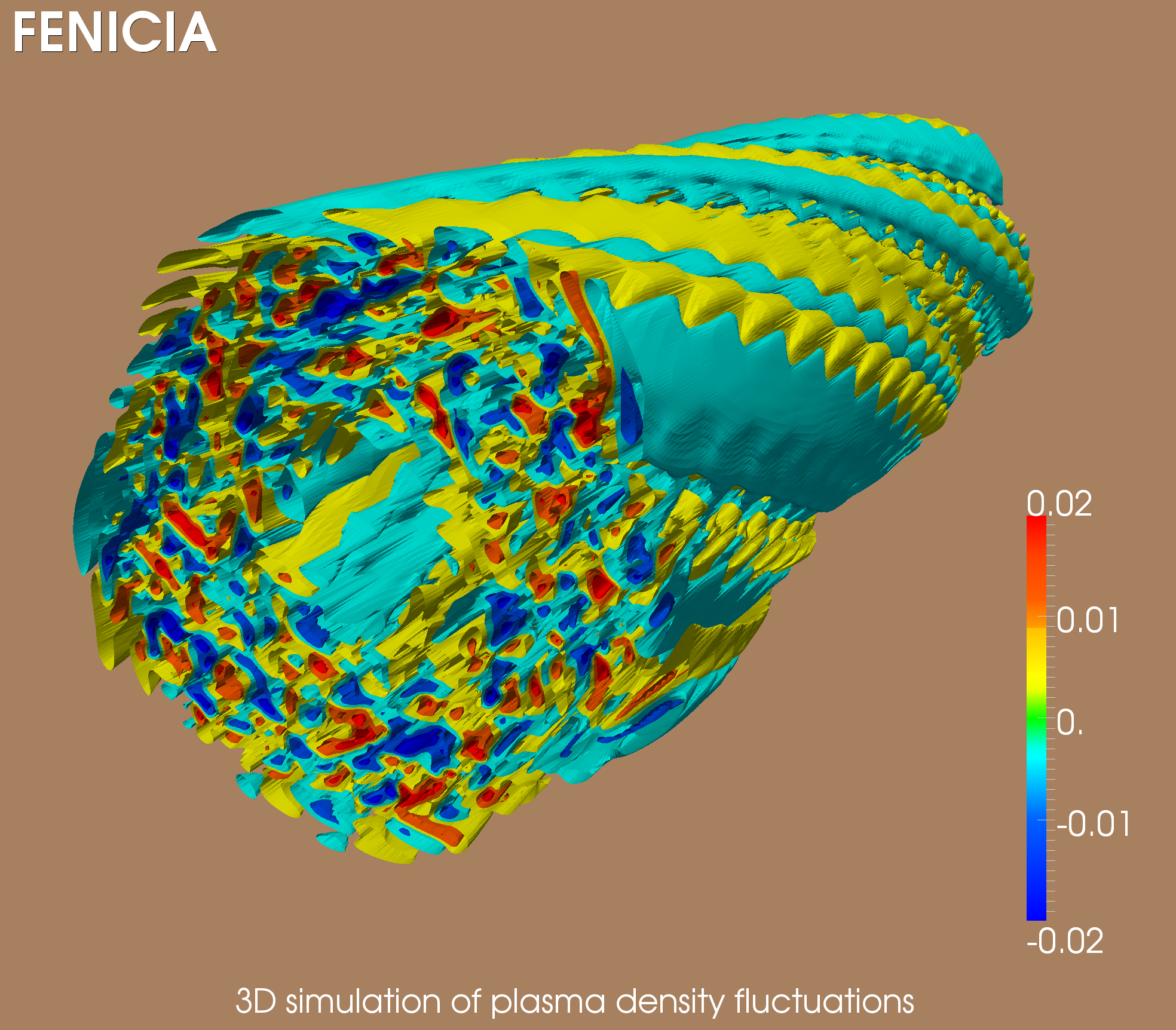}} \hspace*{3ex}
 \hspace*{-0.5cm} \subfloat[t=6]{\includegraphics[width=0.5\linewidth,keepaspectratio,clip]{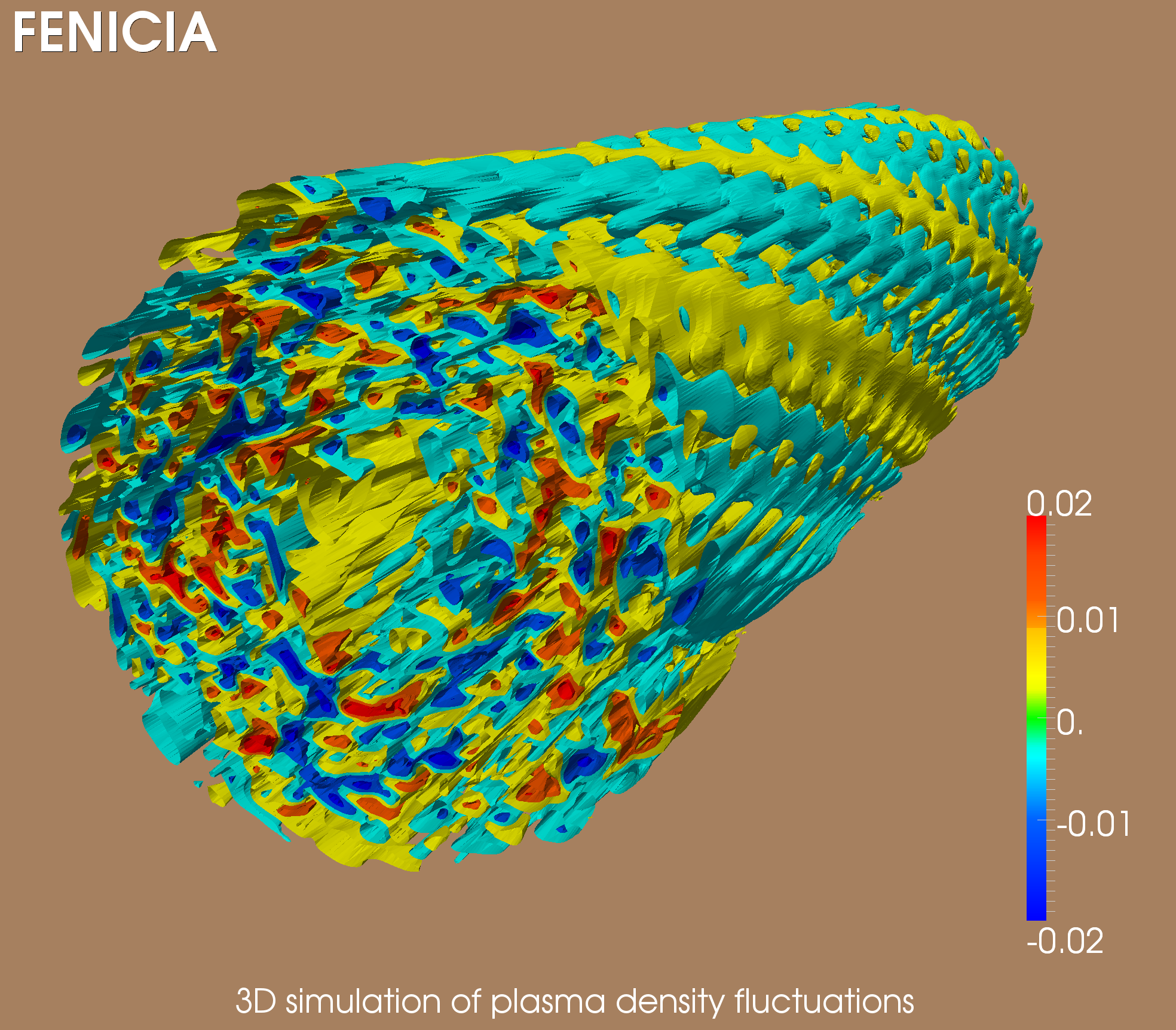}}\\
  \caption{3D snapshots of density fluctuations at different simulation times}
  \label{3D_snapshots}
\end{figure*}

A good amount of information is shown hereafter in figures~\ref{3D_snapshots} of nonlinear simulations carried out using FENICIA with the previously chosen parameters. In the time sequence of the 3D snapshots of density fluctuations, turbulence progressively covers the whole radial domain of the cylinder, but not the center where temperature gradient vanishes for symmetry reason. The entire domain then becomes fully turbulent before the end of the simulation. These figures highlight the fact that turbulent eddies are elongated along the magnetic field lines. When propagating outwards, they encounter regions with larger safety factor values, hence looking more aligned to the $z$ direction.
The reason behind the presence of corrugations in the parallel direction, characterized by finite $k_\parallel$ values, is discussed in what follows.\\

Consider a 3D field $\phi$ in the torus. It can be expressed as a sum of Fourier components along the periodic directions, $\theta$ and $\varphi$ the respective poloidal and toroidal angles, as follows:
\begin{equation}
  \phi(r,\theta,\varphi) = \sum_{m,n} \hat\phi_{m,n}(r,t)\textrm{e}^{i(m\theta+n\varphi)}
\end{equation}
When applied to this field, the dimensionless parallel gradient given by $\nabla_\parallel = \partial_\varphi +q(r)^{-1}\partial_\theta$ in the large aspect ratio limit leads to the following expression:
\begin{equation}
 \nabla_\parallel \phi(r,\theta,\varphi) = \sum_{m,n} i(n+m/q) \hat\phi_{m,n}(r,t) \textrm{e}^{i(m\theta+n\varphi)}
\end{equation}
Thus, the parallel wave vector is simply expressed as $k_\parallel(r) = n + m/q(r)$ and is dependent on the magnetic surface via the safety factor profile $q(r)$. It is apparent that if the field is characterized by almost vanishing parallel gradients $k_\parallel \sim0$, and small scales are to be resolved in a chosen direction (for instance the poloidal direction), then small scales should also be captured in the toroidal direction, i.e: a factor $q^{-1}$ times smaller. Indeed, if $\pm m_{max}$ stands for the highest poloidal wave number of interest for the problem to be addressed, then the highest toroidal wave number should be $n_{max}=\mp m_{max}/q(r)$ in order to ensure that elongated structures in the parallel direction are well described at this scale, namely $k_\parallel(r) = n_{max} + m_{max}/q(r)=0$. In tokamak turbulence, $m_{max}$ is basically constrained by micro-turbulence, which develops at Larmor scales such that $k_\theta\rho_i \sim0.3$. The typical maximal $m$ number of interest for standard situations \footnote{ This may not be the case in regimes with ion transport barriers, where turbulence at electron gyro-radius scales $k_\theta\rho_e \sim0.1$ can become dominant.} is such that $k_{\theta,max}\rho_i \sim2$, so that $m_{max} \sim (r/a)\; \rho_*^{-1}$, which leads to $m_{max}\approx 500$ and $n_{max}\approx 250$ in ITER-like plasmas at mid-radius values. \\

Alternatively, one may compute parallel derivatives using field-aligned coordinates by employing a coarser grid in a chosen direction, while still properly describing the relevant $k_\parallel \sim0$ modes. Although the method proposed in~\cite{Hariri20132419} and implemented in FENICIA is more general than the one discussed in~\cite{Ottaviani20111677}, the reasoning is easier when focusing on the latter one. The parallel gradient can be computed by finite differences along the parallel direction $s$, which reads at second order:
\begin{equation}
 \nabla_\parallel^{FCI} \phi(\rho,\xi,s) \approx
 \frac{\phi(\rho,\xi,s+\Delta s)-\phi(\rho,\xi,s-\Delta s)}{2\Delta s}
\end{equation}
where the superscript $FCI$ stands for field-aligned coordinates, and $\Delta s = \Delta\varphi$ is the increment in the parallel direction. In terms of Fourier modes, an expression as such leads to the following:
\begin{eqnarray}
 \nabla_\parallel^{FCI} \phi(\rho,\xi,s) &=& \frac{1}{2\Delta s}
 \sum_{m,n} \hat\phi_{m,n} (r,t) \nonumber \\
 && \left\{ \textrm{e}^{i[m(\theta-\Delta \varphi/q)+n(\varphi+\Delta \varphi)]} - \textrm{e}^{i[m(\theta+\Delta \varphi/q)+n(\varphi-\Delta \varphi)]} \right\} \nonumber \\
 &=& \sum_{m,n} \hat\phi_{m,n} (r,t)\textrm{e}^{i(m\theta+n\varphi)}
 \frac{\textrm{e}^{i\Delta\varphi(n-m/q)]} -
 \textrm{e}^{-i\Delta\varphi(n-m/q)]}}{2\Delta s} \nonumber \\
 &=& \sum_{m,n} \hat\phi_{m,n} (r,t)\textrm{e}^{i(m\theta+n\varphi)}
 \frac{i\sin[(n-m/q)\Delta\varphi]}{\Delta \varphi}
\end{eqnarray}
It turns out that the effective parallel wave vector then reads $k_\parallel^{FCI} = \sin[(n-m/q)\Delta\varphi] /\Delta \varphi$. In this framework, all modes characterized by $n=m/q + \ell\pi /\Delta\varphi$ (with $\ell\in \mathbb{Z}$) are resonant, in the sense that their effective parallel wave number is vanishing $k_\parallel^{FCI} =0$. In this case, even large $m$ modes can be coupled to ``resonant'' (i.e. such that $k_\parallel^{FCI} =0$) low $n$ modes. As a matter of fact, given a maximum poloidal wave number $m_{max}$ which needs to be resolved, there is no need to go up to $n_{max}=m_{max}/q$ in order to properly account for resonant modes at this small transverse scale: the toroidal $n_\ell$ modes characterized by $n_\ell=m_{max}/q + \ell\pi /\Delta\varphi$ will already do the job.

\begin{figure*}
  \centering
  \includegraphics[width=0.8\linewidth,keepaspectratio,clip]{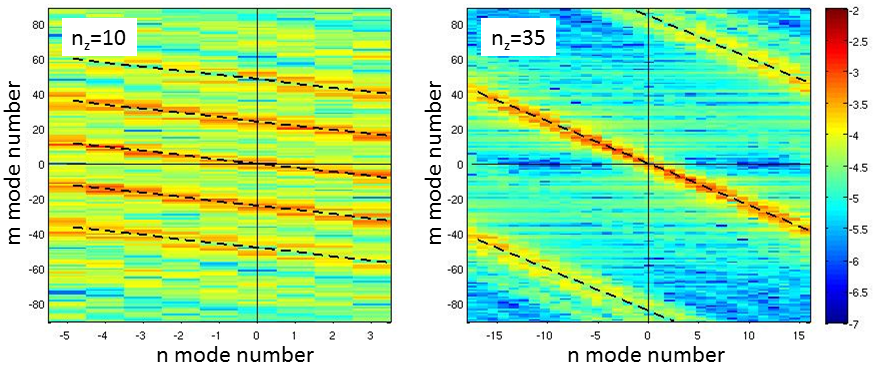}
  \caption{2D Fourier transform of the electric potential $\phi$ at mid-radius and at the end of the simulation runs for two different grid meshes in the axial direction: $n_z=10$ (left) and $n_z=35$ (right). The oblique dash lines satisfy the relation $n_\ell=m/q + \ell\pi /\Delta\varphi$, with $\ell\in \mathbb{N}$ and $\Delta\varphi = 2\pi/n_z$.}
  \label{fig:Fourier2D_scan_nz}
\end{figure*}

\begin{figure}
  \centering
  \includegraphics[width=0.8\linewidth,keepaspectratio,clip]{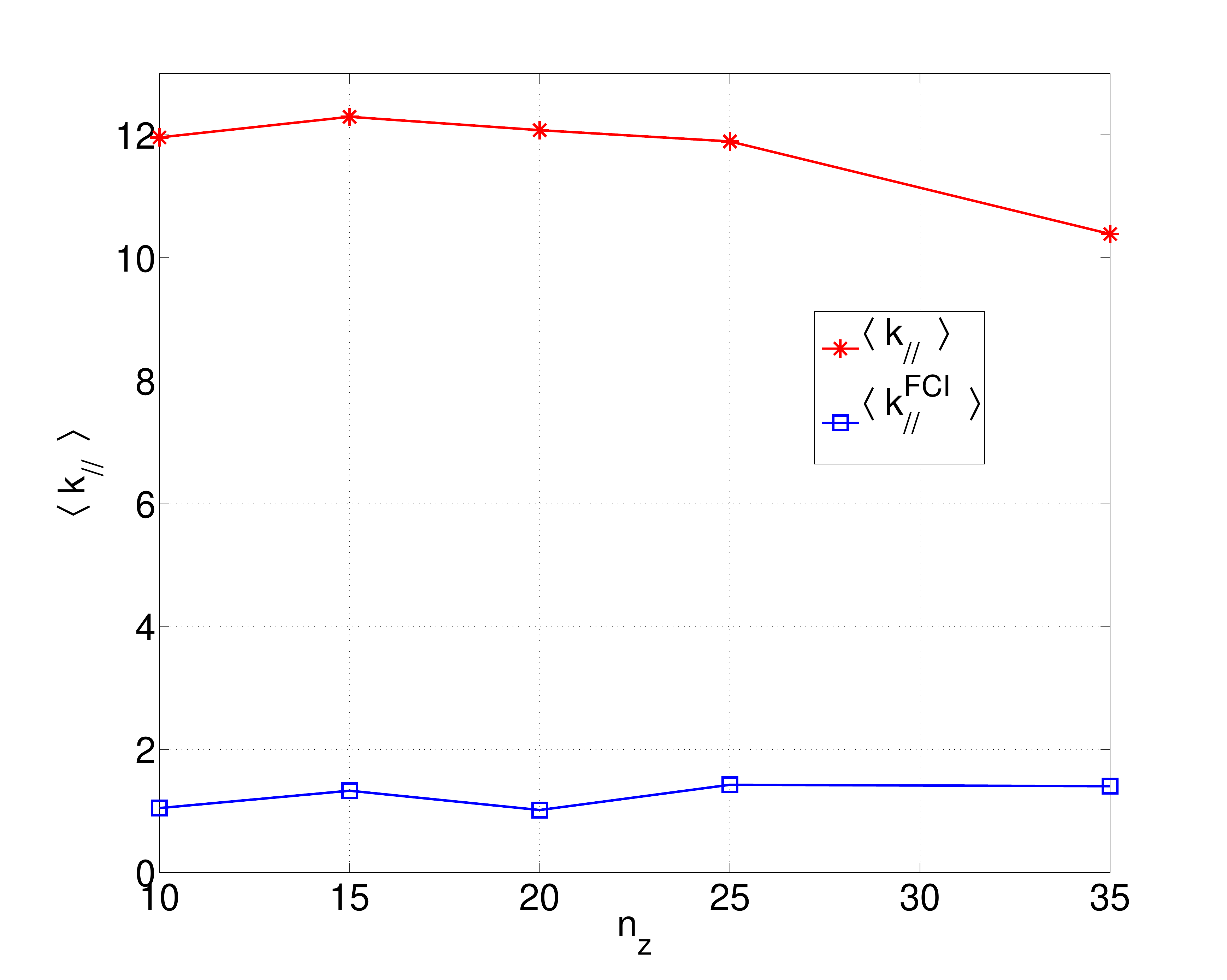}
  \caption{Exact and effective mean parallel wavenumbers (see text) as a function of the number of grid points along $z$.}
  \label{fig:Compare_kparallel_scan_nz}
\end{figure}

\begin{figure}
  \centering
  \includegraphics[width=0.8\linewidth,keepaspectratio,clip]{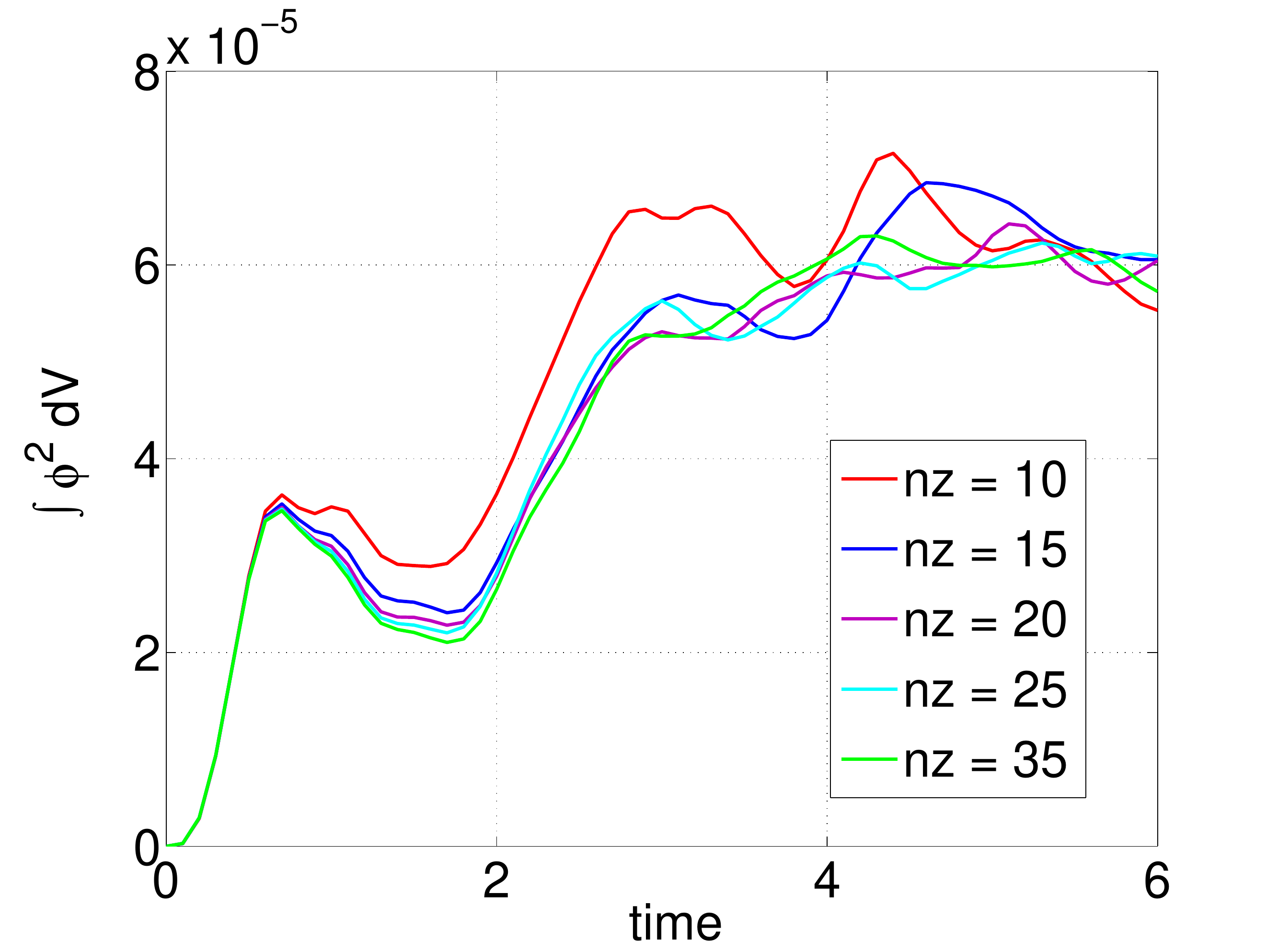}
  \caption{}
  \label{convergence_test}
\end{figure}

This interesting property of field-aligned coordinates is evidenced in figure~\ref{fig:Fourier2D_scan_nz}. We plotted the 2D Fourier transforms of the electric potential $\phi$ at mid-radius at the end of the simulation for different number of grid points in the \textit{z} direction: $n_z=10$ and $n_z=35$. One notices that, in both cases, the spectrum exhibits large amplitude modes (illustrated by the yellow and red colors) outside the exact resonant band $n=-m/q$. This aliasing appears when the number of grid points in $z$ is not sufficient, more precisely when $\Delta\varphi = 2\pi/n_z$ is bigger than half the inverse of the biggest physically relevant wavenumber $n_{max}$ (after the Nyquist-Shannon theorem). In this case, the mean \emph{exact} parallel wavenumber $\langle k_\parallel \rangle$ defined as
\begin{equation}
  \langle k_\parallel \rangle = \left\{
 \frac{\sum_{m,n} (n+m/q)^2\; |\hat\phi_{m,n}(r,t)|^2}
  {\sum_{m,n} |\hat\phi_{m,n}(r,t)|^2} \right\}^{1/2}
\end{equation}
is large. It is equal to $12$ and $10.5$ for these two respective cases (see fig.\ref{fig:Compare_kparallel_scan_nz}). Conversely, the mean \emph{effective} parallel wavenumber $\langle k_\parallel^{FCI} \rangle$, computed from flux-coordinate independent (FCI) field-aligned coordinates and defined as:
\begin{equation}
  \langle k_\parallel^{FCI} \rangle =  \left\{  \frac{\sum_{m,n}
   \left\{ \sin[(n+m/q)\Delta\varphi]/\Delta\varphi \right\}^2\; |\hat\phi_{m,n}(r,t)|^2}
  {\sum_{m,n} |\hat\phi_{m,n}(r,t)|^2} \right\}^{1/2}
\end{equation}
remains small, of the order of $1$ in both cases (cf. fig.\ref{fig:Compare_kparallel_scan_nz}). This is due to the fact that aliasing leads to modes which still satisfy the condition for effective resonance, namely $n_\ell=-m/q + \ell\pi /\Delta\varphi$, with $\ell\in \mathbb{Z}$ and $\Delta\varphi = 2\pi/n_z$ (oblique dash lines on fig.\ref{fig:Fourier2D_scan_nz}). In other words, numerical methods which rely on $(\theta,\varphi)$ coordinates to compute the parallel derivative would interpret those off-diagonal modes as non-resonant ones, with large $k_\parallel$, while field-aligned coordinates are still able to correctly interpret them as effectively resonant ones.
The main result is that the FCI approach permits a coarser mesh in the $\mathit{z}$ direction while still allowing high resolution of the perpendicular spatial dimensions, where the small scales occur, and best representing the physical properties of the model. As Fig.~\ref{convergence_test} shows, values of the square of the electric potential converge starting from $nz = 15$ only. Because the aim is to represent the Physics in the most economical way, it is thus proven that the FCI system is indeed a judicious choice of coordinate transformations best suited for describing the wave dynamics along the field lines and providing us with more efficiency and flexibility in solving anisotropic 3D problems.

\section{Interaction between islands and ITG turbulence}\label{ITGisland}

We have already shown in a previous paper that the FCI approach is well suited to deal with magnetic X-point geometries~\cite{hariri2014flux}.
There, using the FENICIA code, it is shown that analytical sound wave solutions can be recovered with good accuracy, both inside and outside a prescribed static island.
Also, convergence tests were successfully performed for solutions across the separatrix.
In this section, we extend these results to the nonlinear case, showing how turbulence is modified in the presence of a static island.

The main result we focus on deals with the possible impact of turbulence on the temperature gradient inside the island.
This issue is particularly critical in the framework of neoclassical tearing mode (NTM) physics.
Tearing modes often limit the performance of tokamak plasmas, because the magnetic islands which they generate lead to a loss of confinement, and can even lead to a disruption.
NTMs are even more dangerous, since they can grow to a large amplitude because of the amplification effect that the bootstrap current deficit has on an initial ``seed'' magnetic island.
Briefly recalled, the main pieces of the instability mechanism are as follows.
The main drive of the instability comes from the jump in the derivative of $\tilde{\psi}$  across rational surfaces, the so-called $\Delta^\prime$ term.
When created, an island tends to exhibit flat pressure profiles across the O-point, as a result of the strong parallel transport which homogenizes the pressure along magnetic field lines.
In turn, this creates a current hole in the island, since the contribution of the bootstrap current $-$ which is proportional to the pressure gradient and to the collisional regime $-$ is then much reduced.
This actually amplifies the instability drive, the new forcing term being inversely proportional to the island width.
As a matter of fact, for sufficiently small island widths the bootstrap term is the dominant one, so that even in situations when the plasma is stable to the classical tearing mode, the effect of the bootstrap current is to drive it unstable, leading to NTM.

This is where turbulence enters the game.
The island is characterized by an almost vanishing pressure gradient.
As such, it is expected to be turbulence-free, since it is linearly stable.
However, turbulence can penetrate stable regions when excited in neighboring unstable regions.
This is known as turbulence spreading.
When propagating, it obviously modifies the local gradient.
The key questions are then whether turbulence is able to sustain some finite gradient inside an island, and whether this would be sufficient to substantially modify the NTM stability threshold, i.e. the critical size of the seed island required for the growth of NTM.

This section presents preliminary results in this direction, mainly aimed here at illustrating the capabilities of the FCI approach in this magnetic configuration and in the nonlinear regime.
The model equations are close to those discussed in section~\ref{sec:ITGturbulence}, but with the density and parallel velocity equations now being:
\begin{align}
  \partial_t n + [\phi,n] + C_\Vert\nabla_\Vert u &= D_n\nabla_\perp^2 n,   \label{eq:slab-ITG-dndt} \\
  \partial_t u + [\phi,u] + C_\Vert\nabla_\Vert \left\{ \left(1 + \tfrac{1}{\tau}\right)\phi + T_\Vert \right\} &= D_u\nabla_\perp^2 n \label{eq:slab-ITG-dudt}
\end{align}
where $\phi$ is now calculated via the quasi-neutrality equation,
\begin{equation}
  \label{eq:phi-QN}
  n = \phi - \rho_*^2\nabla\phi.
\end{equation}
The lack of the flux-surface average term in Eq.~\eqref{eq:phi-QN} means that the system has weaker zonal flows (ZFs).
This is, however, not important for the goal of this study.
Furthermore, it is known that ZFs in slab ITG (without curvature) would be stronger than the ZFs in the toroidal case, due to the absence of geodesic damping, resulting in a weaker turbulence.
With our choice the two effects are somewhat compensated.

The prescribed magnetic field is given by $\mathbf{B} = \pmb\nabla \times (\psi \mathbf{\hat{z}}) + \mathbf{\hat{z}}$, with $\mathbf{\hat{z}}$ the direction of the magnetic axis.
It accounts for a magnetic island through the expression of the flux function: $\psi =  -(x-x_0)^2/2 + \hat b_x \cos(k_yy)$.
Here, $\hat b_x$ and $k_y$ stand respectively for the magnitude of the radial magnetic field perturbation, which sets the island width, and the island poloidal wavenumber.
The island half-width $w$ is given by $(4\hat{b}_x)^{1/2}$.

The system is initialised with the same parallel and transverse temperature profile, which are allowed to freely evolve in the course of the simulation.
Since Dirichlet conditions are prescribed at both radial boundaries, temperatures remain constant at $x_{min}$ and $x_{max}$.
The forcing then consists of thermal baths.
The rest of the profiles are initialised to zero, except for small perturbations in the density and the potential, calculated via quasi-neutrality.

We use ``ITER-like'' parameters: $R/L_T = 5$, and $\rho_i = 4$mm.
We take the middle of the box to be $r=1$m, and assuming $a=2$m then $\rho_*=1/500$.
The simulation domain is $100\rho_i \times 100\rho_i$ in the poloidal plane, which corresponds to $k_y\sim16$.
This is rather large for ITER magnetic islands, which typically have poloidal modenumbers of $3-5$.
However, this would require a restrictive number of grid-points in the poloidal direction ($\sim6000$).
We use $N_x=N_y=400$ and $N_z=32$ points in the radial, poloidal and toroidal directions, respectively.

\begin{figure}
  \centering
  \includegraphics[width=0.8\linewidth]{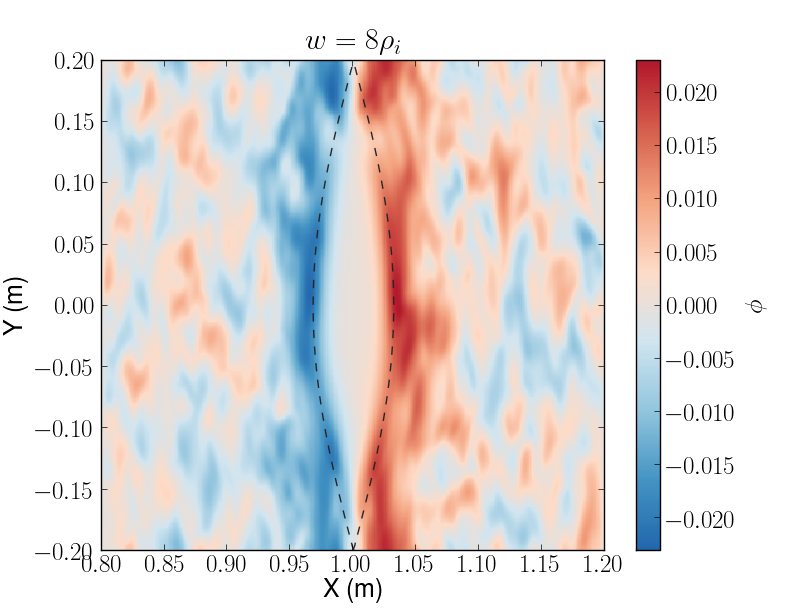}
  \caption{2D snapshot of the electrostatic potential for an island with $w=8\rho_i$.}
  \label{fig:rho8-2Dpot}
\end{figure}

\begin{figure}
  \centering
  \includegraphics[width=0.8\linewidth]{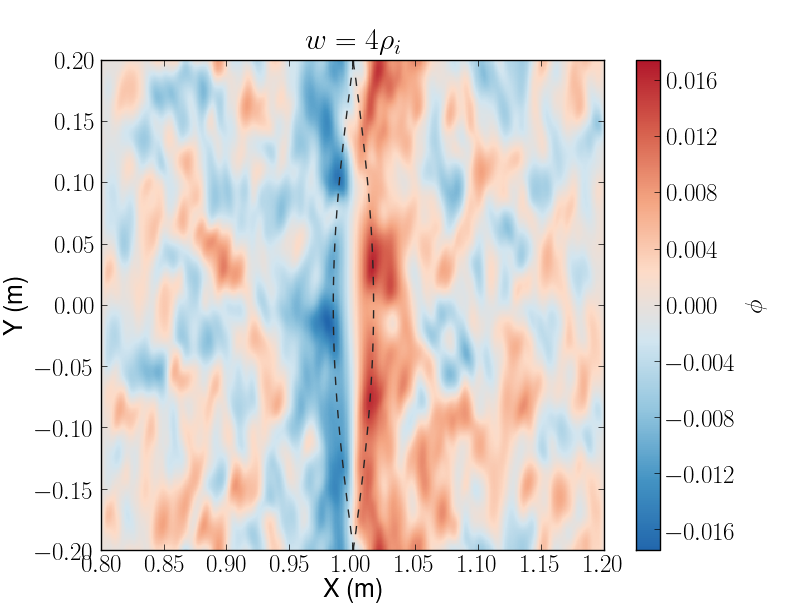}
  \caption{2D snapshot of the electrostatic potential for an island with $w=4\rho_i$.}
  \label{fig:rho4-2Dpot}
\end{figure}

The simulation is run for one Bohm time without an island, until turbulence has developed and saturated, at which point the simulation is restarted and the island is turned on.
Due to the relative low cost of these simulations, we are able to run them for 5--6 Bohm times. Figures~\ref{fig:rho8-2Dpot} and~\ref{fig:rho4-2Dpot} show 2D snapshots of the electrostatic potential at 2 Bohm times for islands of sizes $w=8\rho_i$ and $w=4\rho_i$, respectively. We see that turbulence spreads inside the island of width $w=4\rho_i$, but it does not penetrate the one with $w=8\rho_i$.

Simulations without turbulence were also performed.
This is done by keeping just the $T_\perp$ equation and dropping the terms containing $\phi$.
The results from the turbulence-free simulations are in agreement with \cite{fitzpatrick1995helical}: there is a large island limit ($w\gtrsim8\rho_i$) where parallel transport dominates and the temperature profile is completely flattened, and a small-island limit ($w\lesssim2\rho_i$) where the perpendicular transport dominates and the temperature profile is almost completely unaffected.
In-between these two limits is the intermediate region, where parallel and perpendicular transports are comparable, and the temperature profile is not completely flattened in this region; instead, the degree of flattening is determined by the island width.

When turbulence is present, the large- and small-island limits are recovered.

\begin{figure}
  \centering
  \includegraphics[width=0.7\linewidth]{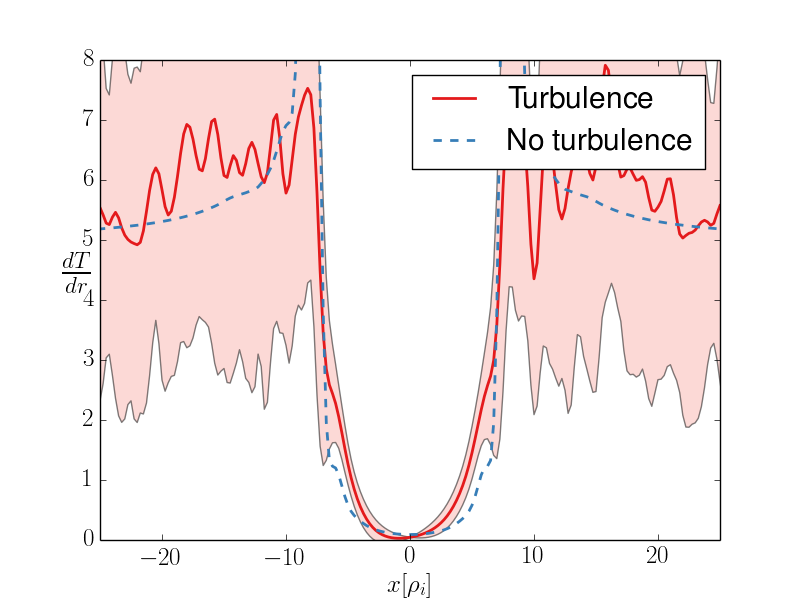}
  \caption{Perpendicular temperature gradient profiles through the O-point with (red) and without (blue) turbulence for an island with $w=8\rho_i$.}
  \label{fig:rho8-dTdr}
\end{figure}

\begin{figure}
  \centering
  \includegraphics[width=0.7\linewidth]{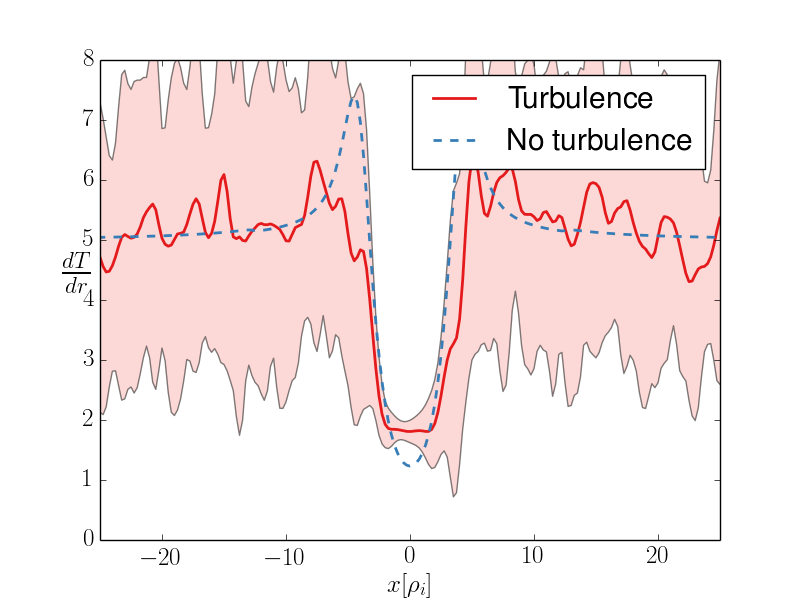}
  \caption{Perpendicular temperature gradient profiles through the O-point with (red) and without (blue) turbulence for an island with $w=4\rho_i$.}
  \label{fig:rho4-dTdr}
\end{figure}

Figure~\ref{fig:rho8-dTdr} illustrates how the temperature profile is completely flattened inside the island, although it is modified outside the separatrix.
In the intermediate region, however, the temperature profile inside the island is modified by the presence of turbulence, see Fig.~\ref{fig:rho4-dTdr}.
Turbulence is able to penetrate the interior of the island and sustain an increased temperature gradient, over that found without turbulence.
This modification is not well reproduced by using the effective diffusivity, calculated without an island.
Indeed, saturation of the temperature gradient at the O-point occurs on turbulent transport timescales, rather than on collisional ones.
This indicates that one must take into consideration characteristics of the particular turbulence when determining the critical width of seed islands for NTMs.

A more rigorous study of the intermediate island-width region will appear in a future work.

\section{Conclusions}
The Flux-Coordinate independent approach presented in~\cite{Hariri20132419, hariri2014flux} is based on two concepts. The first is using an arbitrary mesh not based on magnetic flux coordinates; the second is computing the parallel gradient operator by tracing the magnetic field lines from one poloidal plane to the next and interpolating at end points. In this paper, the FCI approach is formulated to handle 3D magnetic configurations. The existence of a coordinate transformation is guaranteed only when the system possesses 'good' magnetic surfaces identified by given values of $\psi^*$. In the non-integrable case, such as in stochastic magnetic fields, the FCI approach must be formulated through a direct local integration of the field line equations and interpolation at end points. The FCI approach can thus be used for 3D magnetic field configurations, including stochastic ones.\\

In the second part of the paper, fully nonlinear simulations of a 3-dimensional 4-field ITG model are presented using the FCI formulation for axisymmetric geometry. For the first time, results show the robustness of FCI in a fully turbulent regime. Furthermore, using ITER-like parameters, we simulate the effect of the presence of a static magnetic island on ITG turbulence. First results show that the island affects the temperature gradient immediately outside the separatrix, increasing turbulent fluctuations. Turbulence can also spread inside the island, reducing the degree of flattening. This is qualitatively in agreement with the results found in~\cite{poli2002reduction, la2006cross, poli2009behaviour, hornsby2010nonlinear}. This further confirms the applicability of the FCI approach to a wide range of problems.

\def\newblock{\hskip .11em plus .33em minus .07em}


\end{document}